\documentclass[11pt,twoside]{article}

\usepackage[activeacute,spanish]{babel}
\usepackage{baaa-cas}
\usepackage{graphicx}

\usepackage[T1]{fontenc} 
\usepackage{natbib}         
\usepackage{xcolor}
\usepackage{latexsym}
\usepackage{verbatim}
\usepackage{amssymb}
\begin{document}
\def\tablename{Table}%
\markboth{P. Cataldi et al.}%
{Galaxy sizes in the early Universe}
\newcommand{\pedro}[1]{\textbf{\textcolor{blue}{#1}}}

\pagestyle{myheadings}
%
%
\vspace*{0.3cm}
\parindent 0pt{Plenary/Contributed talk}

\title{\centering{Galaxy sizes in the early Universe}}

\author{P. Cataldi $^{1}$,  S. Pedrosa$^{1}$,  L.J. Pellizza$^{1}$, D. Ceverino$^{2,3}$, L.A. Bignone$^{1}$}

\affil{%
(1) Instituto de Astronom{\'\i}a y F{\'\i}sica del Espacio, CONICET--UBA, Argentina \\
(2) Departamento de Física Teórica, Universidad Autónoma de Madrid, Madrid, Spain \\
(3) CIAFF, Facultad de Ciencias, Universidad Autónoma de Madrid, 28049 Madrid, Spain \\
}

\begin{abstract}
We present preliminary results of a study of galaxy sizes at redshift $z \gtrsim 5$ using the high-resolution, zoom-in \textsc{FirstLight} simulations. We obtain a redshift-independent, chevron-shaped stellar mass--size relation with a peak at $M^{\star} \approx 10^{8.5}~\mathrm{M}_\odot$. This relation is shaped by the individual evolution of galaxies, which undergo a phase of contraction at this mass scale triggered by a strong central starburst.
\end{abstract}

\section{Introduction}

Sizes are fundamental proxies for tracing galaxy formation history and evolutionary processes, some of which (chemical evolution, energy feedback) are driven by massive stars \citep[][and references therein]{Hopkins2023}. Recent JWST observations have found compact (effective radii of the order of a few hundred parsecs), red, massive (stellar masses $M_\star \gtrsim 10^{10}~\mathrm{M}_\odot$) galaxies at redshifts $z > 7$. These also show high star formation rate densities and evidence of inside-out growth \citep[e.g.][]{Morishita2024}. Their origin is still under debate, as it is the nature of the stellar mass--size relation at high $z$. Some authors claim that this relation is inverted \citep[e.g.][]{Baggen2023}, whereas others argue that it remains normal \citep[e.g.][]{Morishita2024}. This highlights some open questions on the mechanisms driving galaxy formation in the early Universe.

Several simulations with varying baryonic physics and resolution \citep{Ni2020,Lovell2021,Kannan2022} reveal an intrinsic mass--size relation with a hump around $M_\star \sim 10^{8}$–$10^{9} \, \mathrm{M}\odot$. A characteristic \textit{turn-on} mass marks the transition from low-mass galaxies, in which feedback-driven outflows dominate and sizes grow, to massive systems shaped by disk instabilities, angular momentum loss, and subsequent compaction \citep{Lapiner2024}. However, other simulations \citep{Shen2024} find no compaction phase for low-mass galaxies ($M\star \sim 10^{7-9} \, \mathrm{M}_\odot$) at $z>3$. These rather exhibit rapid size fluctuations driven by the interplay between feedback outflows and cold inflows. This preserves the overall size–mass relation, increasing its scatter rather than reversing its slope. 

Thus, whether a true hump or inversion exists remains under debate. Given the conflicting results from both observations and simulations regarding the high-redshift size–mass relation, a comprehensive investigation is required to clarify the issue and underpin the underlying physical processes. A step forward can be made by using zoom-in simulations, especially devised to explore galaxy evolution at Cosmic Dawn, such as the \textsc{FirstLight} suite \citep{Ceverino2017}. Their high resolution allows to precisely trace the mass--size relation through a large range of masses. Here we present some preliminary results of such a study; the full investigation can be found in \citet{Cataldi2025}.

\section{Simulations and results}
\label{procedure}

We use a subsample from the \textsc{FirstLight} cosmological zoom-in simulations, which achieve $\sim10 \, \mathrm{pc}$ spatial resolution and include both thermal and radiative feedback  \citep[for details see][]{Ceverino2017}. The sample comprises 169 galaxies with stellar masses spanning from $\sim10^6$ to $10^{10.5}~\mathrm{M_\odot}$, tracked across 45 snapshots between $z=9$ and $5.25$. It enables us to study the growth of galaxies during Cosmic Dawn, and the possible inversion of the size–mass relation.

\begin{figure}
\begin{center}
\includegraphics[angle=0,height=7.7cm]{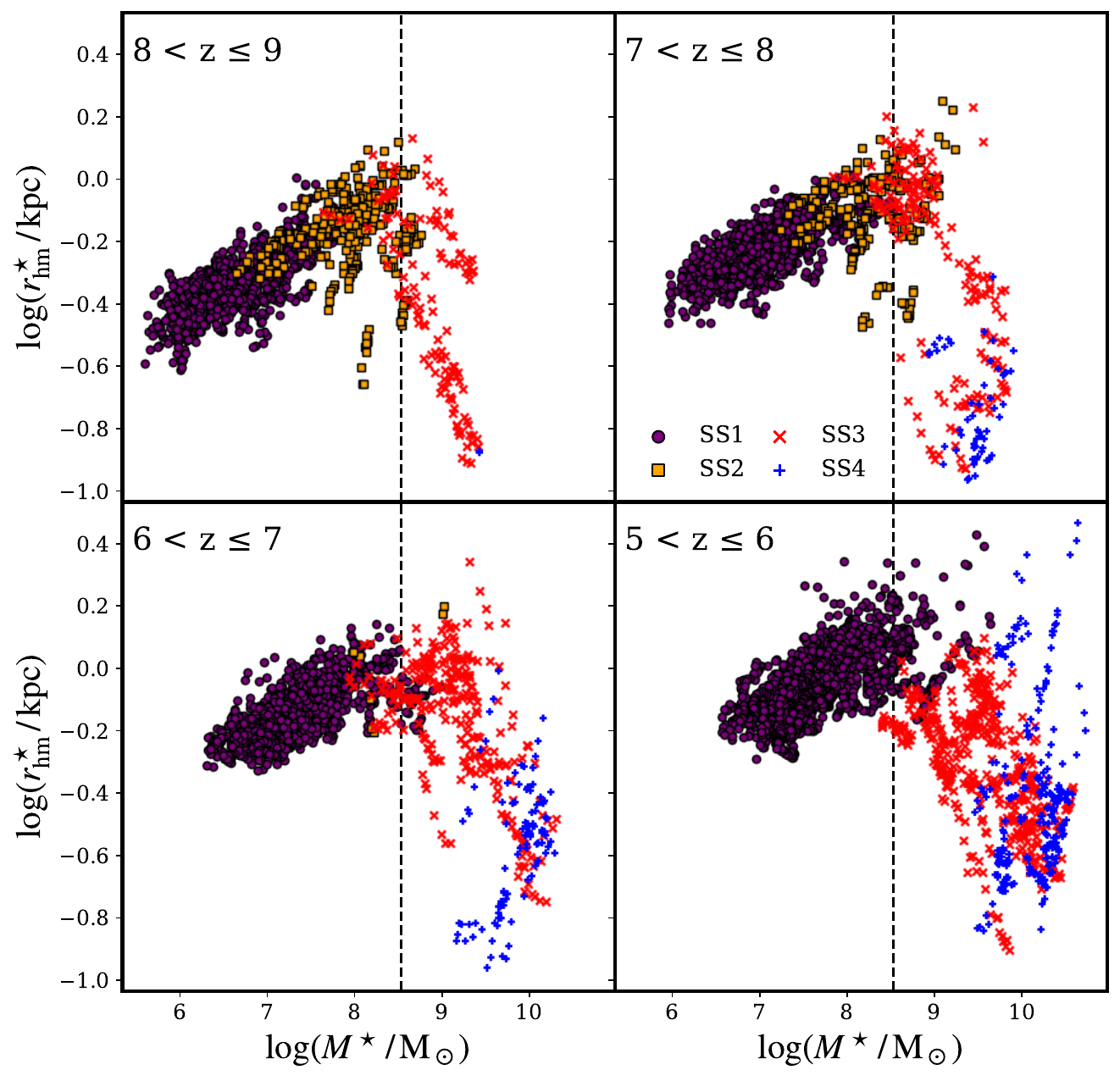}
\caption{Stellar mass--size relation across redshift as a result of evolutionary stages: expanding galaxies (purple dots) lie in the ascending branch, at the top of which we find objects showing the onset of contraction (orange squares). The descending branch is populated by contracting galaxies (red crosses), whereas at its bottom, the systems show an incipient re-expansion (blue plus symbols). The dashed vertical line marks the \textit{turn-on} stellar mass, $M^\star_\mathrm{on} = 10^{8.5}~\mathrm{M}_\odot$.}
\label{fig:fig1}
\end{center}
\end{figure}

Fig.~\ref{fig:fig1} shows the relation between the stellar mass $M^\star$ and the galaxy size (measured by the half-mass radius $r^\star_\mathrm{hm}$) for the sample. A chevron shape is evident at all redshifts, with a tight ascending (\textit{normal}) branch in which galaxy size increases with its mass, and a descending one with larger scatter toward lower redshifts. The peak separating both branches occurs at $M^\star_\mathrm{on} \approx 10^{8.5}~\mathrm{M}_\odot$,  which we call \textit{turn-on} mass. The fact that the loci of both branches are roughly constant with redshift suggests that the relation follows the evolutionary path of galaxies. We confirmed this by analysing the individual mass--size evolution of each galaxy in our sample. We divided our sample into four subsamples characterised by distinct evolutionary patterns: (a) galaxies undergoing continuous expansion over the redshift interval studied (purple dots in Fig.~\ref{fig:fig1}), (b) galaxies with an initial expansion phase prior to compaction (orange squares), (c) galaxies undergoing a compaction phase (red crosses), and (d) galaxies with a re-expansion phase (blue plus symbols). We interpret these as different stages of a single evolutionary sequence in which galaxies expand, then contract, and subsequently re-expand. The existence of the characteristic mass $M^\star_\mathrm{on}$ suggests that stellar mass is the main driver of the compaction process. A similar \textit{turn-off} mass, $M^\star_\mathrm{off} \approx 10^{9.6}~\mathrm{M}_\odot$, appears to mark the point where galaxies start to re-expand, although the small sample of galaxies with masses above $M^\star_\mathrm{off}$ prevents the extraction of meaningful results on this phase. Variations in the critical masses likely reflect differences in star formation histories and environmental or merger effects.

\begin{figure}  
\begin{center}
\includegraphics[height=7.5cm]{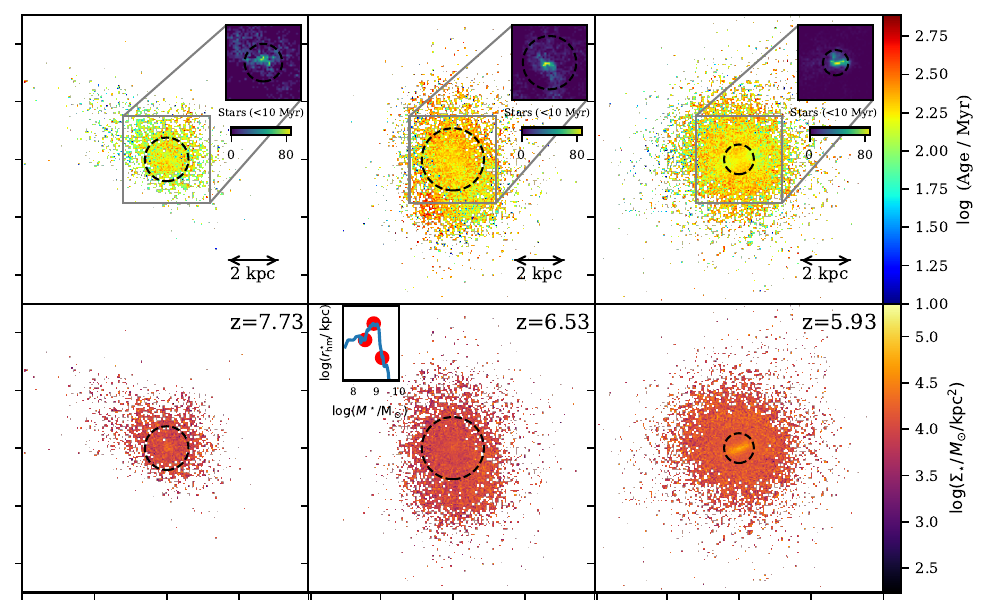}
\caption{Edge-on view of a galaxy at its maximum size (centre panels), and $\sim100~\mathrm{Myr}$ before (left panels) / after (right panels) attaining it. Colours indicate the stellar ages (top) and surface densities (bottom). The black dashed circle marks the half-mass radius $r^{\star}_\mathrm{hm}$. Inset panels highlight newly formed stars within $r^{\star}_{\mathrm{hm}}$ (top) and the mass--size relation, with the red dots marking the loci of the three snapshots (bottom).}
\label{fig:fig3}
\end{center}
\end{figure}	

Fig.~\ref{fig:fig3} illustrates the representative case of a galaxy that undergoes rapid compaction. The sequence reveals the transition from an early, spatially extended star-forming phase characteristic of systems growing inside-out, to a compact distribution of stars concentrated at the galaxy centre. This suggests that contraction is driven by the triggering of a strong starburst in the central regions of galaxies. As the half-mass radius depends on the distribution of the stellar mass, a large mass of stars must be formed in these regions to stop and finally reverse its increase.

\section{Conclusions}
\label{discussion}

We identify a population of compact galaxies formed at high redshift via a mechanism that temporarily stops normal inside-out growth, rapidly shrinking their stellar distributions to sizes comparable to those observed. This process produces a chevron-shaped stellar mass--size relations with a peak mass $M^\star_\mathrm{on} \approx 10^{8.5},\mathrm{M}_\odot$, independent of redshift, indicating that stellar mass is its main driver. Growth then resumes after this compaction, near a characteristic mass $M^\star _\mathrm{off} \simeq 10^{9.6},\mathrm{M}_\odot$. Our results suggest that the origin of compaction is the triggering of a strong starburst at the central regions of galaxies, which supports a wet compaction scenario \citep{Dekel2014}, which may be driven by inflows of cold gas from the outskirts. Future work will investigate the underlying physical mechanisms responsible for sustaining and regulating this process.

\acknowledgements Authors acknowledge support from ANPCyT through PICT 2020-00582, and the European Union’s Horizon 2020 Research and Innovation Programme (Marie Skłodowska-Curie grant agreement No 734374 / LACEGAL). DC is supported by research grants PID2021-122603NB-C21 funded by the Ministerio de Ciencia, Innovación y Universidades (MICIU/FEDER) and CNS2024-154550 funded by MICIU/AEI/10.13039/501100011033. The authors gratefully acknowledge the Gauss Center for Supercomputing for funding this project by providing computing time on the GCS Supercomputer SuperMUC at Leibniz Supercomputing Center (Project ID: pr92za), computer resources at MareNostrum, and technical support by the Barcelona Supercomputing Center (RES-AECT-2020-3-0019). 

\bibliographystyle{aa}
\bibliography{bibliography}

\end{document}